# Tunable altermagnetism via inter-chain engineering in parallel-assembled atomic chains

Deping Guo[1,2 #], Canbo Zong[2,3 #], Weihan Zhang[2,3], Cong Wang[2,3 *], Junwei Liu[4*] and Wei Ji[2,3*]

[1]*College of Physics and Electronic Engineering, Center for Computational Sciences, Sichuan Normal University, Chengdu, 610101, China*

[2]*Beijing Key Laboratory of Optoelectronic Functional Materials & Micro-Nano Devices, School of Physics, Renmin University of China, Beijing 100872, China*

[3] *Key Laboratory of Quantum State Construction and Manipulation (Ministry of Education), Renmin University of China, Beijing, 100872, China*

[4]*Department of Physics, Hong Kong University of Science and technology, Hong Kong 999077, China.*

[*]*Corresponding authors.* Email: wcphys@ruc.edu.cn (C.W.) liuj@ust.hk (J.L), wji@ruc.edu.cn (W.J.)

[#] *These authors contributed equally to this work*

**Altermagnetism has recently drawn considerable attention in three- and two-dimensional materials. Here, we extend this concept to quasi-one-dimensional (Q1D) monolayers assembled from single-atomic magnetic chains. Through systematically examining nine types of structures, two stacking orders, and intra-/inter-chain magnetic couplings, we identify four out of thirty promising structural prototypes for hosting altermagnetism, which yields 192 potential monolayer materials. We further confirm eight thermodynamically stable Q1D monolayers via high-throughput calculations. Using symmetry analysis and first-principles calculations, we find that the existence of altermagnetism is determined by the type of inter-chain magnetic coupling and predict three intrinsic altermagnets, $CrBr_3$, $VBr_3$, and $MnBr_3$, due to their ferromagnetic inter-chain couplings and five extrinsic ones, $CrF_3$, $CrCl_3$, $CrI_3$, $FeCl_3$, and $CoTe_3$, ascribed to their neglectable or antiferromagnetic inter-chain couplings. Moreover, the inter-chain magnetic coupling here is highly tunable by manipulating the inter-chain spacing, leading to experimentally feasible transitions between altermagnetic and nodal-line semiconducting states. In addition, applying external electric fields can further modulate the spin splitting. Our findings establish a highly tunable family of Q1D altermagnets, offering fundamental insights into the intricate relationship**

**between geometry, electronic structure, and magnetism. These discoveries hold significant promises for experimental realization and future spintronic applications.**

Recent advancements in magnetic and spin groups uncovered many unconventional antiferromagnetic (AFM) materials exhibiting zero net magnetization in real space and spin-split bands with crystal symmetry paired spin-momentum locking in momentum space even without spin-orbit coupling [1–6] and can be described by the spin group [7–12], a collinear subset of which is termed altermagnets [13–16]. The magnetic unit cell of an altermagnet comprises an even number of magnetic atoms that form two magnetic sublattices with antiparallel spins, which are related by crystallographic rotation or mirror operations, rather than inversion or translation symmetry operations, leading to the breaking of joint parity and time (P-T) symmetry or time reversal and translation symmetry ($T\tau$) [13,14]. Altermagnets combine the advantages of both ferromagnets (spin-polarized bands) and antiferromagnets (zero net magnetization), thereby mitigating stray magnetic fields related issues. As a result, they exhibit a range of novel properties [4,6,17–33], such as the giant piezomagnetism [6], noncollinear spin current [6,17], and anomalous Hall effect [29,30]. To date, more than 200 intrinsic three-dimensional altermagnetic materials have been theoretically predicted, with some experimentally confirmed [15,16] . However, the prediction of intrinsic two-dimensional (2D) altermagnets remains comparatively limited [6,17–19,34–37], primarily ascribed to stringent symmetry constraints intrinsic to 2D materials. Though these inherent geometric limitations can be transcended through artificial engineering [38–46], including the introduction of twisting angle [38] and application of external electric fields [46] , the exploration of their tunability is still in its infancy.

One-dimensional (1D) materials have largely been overlooked as potential altermagnets, even including artificially engineered structures, due to their restrictive symmetry operations. Recent experimental breakthroughs have demonstrated the synthesis of single-atomic magnetic chains of $CrCl_3$ on surface [47] or encapsulated within carbon nanotubes [48,49], and their self-assembled ribbons [47]. In these ribbons, while strong covalent bonds connect atoms within individual chains, van der Waals (vdW) interactions predominantly govern inter-chain couplings. This anisotropic nature of interactions highlights the feasibility of artificially arranging these chains into Q1D monolayers with tunable inter-chain spacing and stacking order. The highly adjustable inter-chain geometry differentiates these Q1D monolayers from conventional 2D monolayers, enabling precise control over their inter-chain magnetic couplings [50–52]. Such controllable inter-chain magnetism can be strategically tuned to meet the symmetry requirements necessary for altermagnetism, presenting distinctive opportunities to extend the scope of monolayer altermagnets and to explore novel altermagnetic properties.

Here, we employed a self-assembly strategy to construct Q1D monolayer configurations from $XY_n$ 1D single-atomic magnetic chains. By systematically exploring chemical stoichiometric ratios, structural phases, stacking orders, and both inter-chain and intra-chain magnetic couplings, we identified four altermagnetic structural prototypes among 30 candidates. Expanding on these prototypes, we generated 192 monolayer materials and screened eight dynamically stable monolayers through high-throughput calculations. To verify the presence of altermagnetism in these materials, we calculated their band structures and magnetic exchange parameters. Furthermore, we investigated the effects of inter-chain spacing and external electric fields on altermagnetism of Q1D monolayers.

Our density functional theory (DFT) calculations were carried out using the generalized gradient approximation for the exchange-correlation potential [53], the projector augmented wave method [54] and a plane-wave basis set as implemented in the Vienna ab-initio simulation package (VASP) [55,56]. All calculations, the Grimme's D3 form vdW correction was applied to the Perdew Burke Ernzerhof (PBE) exchange functional (PBE-D3) [57]. Kinetic energy cut-off of 700 eV and 500 eV for the plane wave basis set were used in structural relaxations and electronic calculations, respectively. All atomic positions and lattices were fully relaxed until the residual force per atom was less than 0.001 eV/Å. 2×2×1 supercells are used for the calculations of structural relaxation and total energy. An 8×2×1 *k*-mesh was adopted to sample the Brillouin zone of the 2×2×1 monolayer supercells. A vacuum layer, over 15 Å in thickness, was used to reduce interactions among image slabs. On-site Coulomb interactions on *d* orbitals of the Cr (U= 3.9 eV, J=1.1 eV) for $CrF_3$, $CrCl_3$, $CrBr_3$ and $CrI_3$, V (U= 3.0 eV) for $VBr_3$, Mn (U= 4.9 eV) for $MnBr_3$, Co (U= 4.9 eV) for $CoTe_3$ and Fe (U= 3.9 eV) for $FeCl_3$ were considered using a DFT+U method [58]. The values of *U* were determined using a linear response method [59] based on the 2D structure with the same chemical formula to the Q1D chains and their robustness has been tested (Fig.S1 [60]).

Nine structural phases have been experimentally observed and/or theoretical predicted in binary 1D single-atomic magnetic chains, covering four chemical stoichiometric ratios (1:1, 1:2, 1:3, and 1:4) between transition metal (X) and non-metal (Y) elements [61–64]. Given that altermagnets represent a subset of AFMs requiring zero net magnetization, three categories of magnetic ordering in Q1D monolayers can potentially exhibit altermagnetism: (1) intra- and inter-chain AFM, (2) intra-chain AFM with inter-chain FM, and (3) intra-chain FM with inter-chain AFM. In the third category, opposite spins are related by a fractional translational symmetry operation, which excludes the possibility of altermagnetism. Therefore, our following analysis specifically targets the first two categories, where magnetic chains exhibit intra-chain AFM ordering. These intra-chain AFM coupled chains can self-assemble into Q1D monolayers in either AA or AB stacking, with the inter-chain magnetism being either FM or AFM. Considering variations across stoichiometric ratios, structural phases, inter-chain stacking orders, and magnetic ordering, we systematically 30 Q1D monolayer prototypes, as summarized in Fig. 1 and Figs. S2 to S6 of the Supplemental Materials [60].

(a)

| Configuration | | AA | | AB | |
|---|---|---|---|---|---|
| Ratio | Phase | FM | AFM | FM | AFM |
| XY | α | × | × | × | × |
| $XY_2$ | α | × | × | / | |
| | β | × | × | × | × |
| | γ | √ | × | × | √ |
| $XY_3$ | α | × | × | / | |
| | β | √ | × | × | √ |
| | γ | × | × | / | |
| $XY_4$ | α | × | × | × | × |
| | β | × | × | × | × |

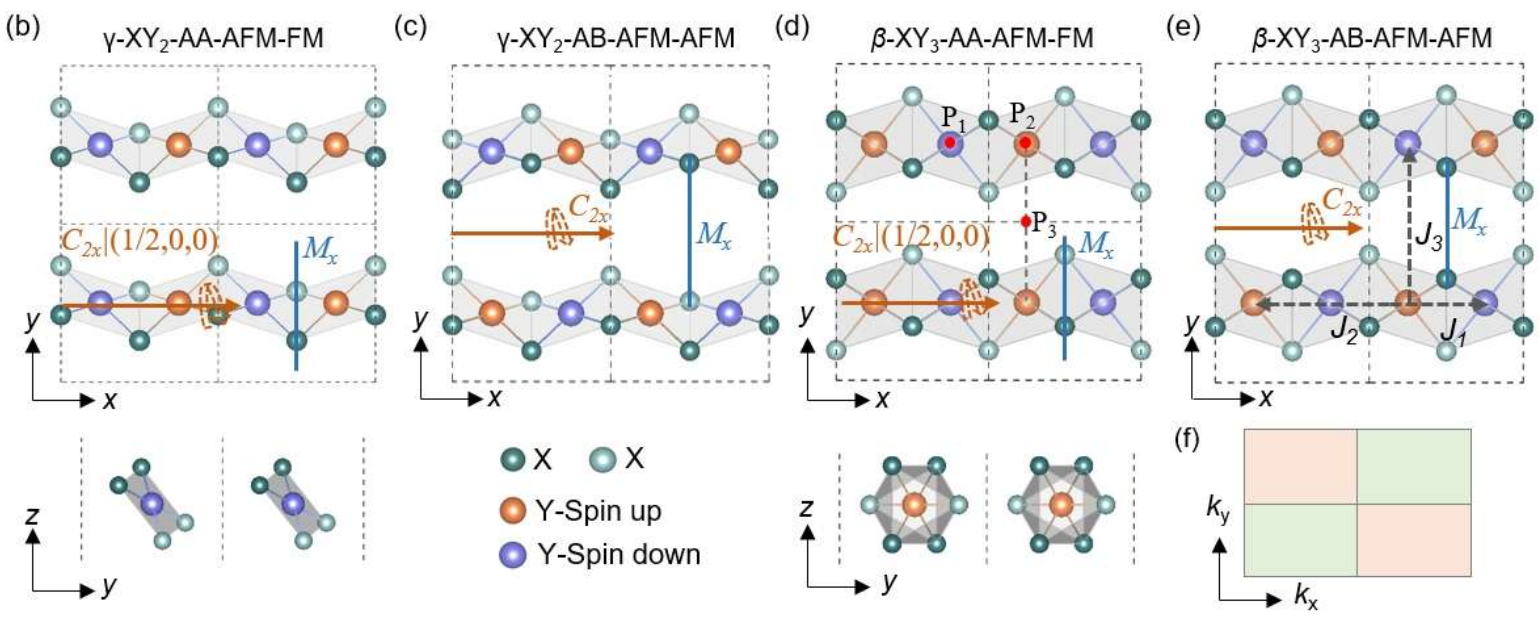

FIG. 1. (a) Summary of the emergence of altermagnetism in 1D magnetic chains with different stoichiometric ratios under AA and AB stacking configurations. FM and AFM represent inter-chain magnetic ordering. Symbol '×' indicates the absence of altermagnetism, while '√' signifies its emergence. Symbol '/' represents the absence of the AB stacking configuration. Top (upper panel) and side (lower panel) views of the AA-stacked (b) and AB-stacked (c) $\gamma$-phase-$XY_2$ (X=transition-metal, Y=chalcogen/halogen-atom), and AA-stacked (d) and AB-stacked (e) $\beta$-phase-$XY_3$ monolayers. Orange arrows and blue lines illustrate symmetry operations $C_{2x}$ and $M_x$ that connect the sublattices with opposite spins. Red dots $P_1$ to $P_3$ marked in panel (d) indicate structural inversion centers. Orange and blue spheres represent magnetic atoms with up and down majority spins, respectively. $J_1$, $J_2$, and $J_3$ marked in panel (e) represent spin-exchange parameters for the nearest-, second-nearest-, and third-nearest-neighbors, respectively. (f) Diagram of spin-splitting symmetry in the Brillouin zone.

Among these 30 prototypes examined, four satisfy the symmetry requirements for altermagnetism, characterized by broken P-T symmetry, as indicated by check marks in Fig. 1(a). These include two $\gamma$-phase $XY_2$ monolayers [prototype-1, Fig. 1(b) and prototype-2, Fig. 1(c)] and two $\beta$-phase $XY_3$ monolayers [prototype-3, Fig. 1(d) and prototype-4, Fig. 1(e)]. Prototypes-1 and -3 adopt AA stacked inter-chain FM configurations [Figs. 1(b) and 1(c)] while the remaining two are in AB stacked inter-chain AFM configurations [Figs. 1(d) and 1(e)]. Each prototype structure maintains an inversion symmetry but lacks the P-T symmetry, which enabling the emergence of altermagnetism. Their sublattices with opposite spins are related by the $C_{2x}$ (orange) and $M_x$ (blue) symmetries. As a result, the spin splitting distribution maps in the Brillouin zone simultaneously reflect these two symmetries, giving rise to an overall *d*-wave-like ($C_{4z}$) symmetry [Fig. 1(f)]. Taking prototype-3 [Fig. 1(d)] as an illustrative example, three inversion centers, $P_1$, $P_2$ and $P_3$ (marked by red dots), can be identified. Independent of inter-chain magnetism, the inter-chain inversion centers $P_1$ and $P_2$ link atoms sharing the same majority spin, thereby inherently breaking the P-T symmetry. Under inter-chain FM couplings, the inter-chain inversion center $P_3$ also links atoms with the same spin, further strengthening the broken P-T symmetry and thereby facilitating the altermagnetism. Conversely, tuning the inter-chain FM coupling to AFM restores the P-T symmetry, thereby preventing the emergence of altermagnetism.

By realizing all four prototypes through six fourth-period transition-metals and eight chalcogen or halogen non-metal elements, we constructed 192 Q1D monolayers as potential altermagnet candidates, see Fig. 2(a) blue block. Among them, 50 monolayers consist of single-atomic chains with intra-chain AFM ground states (Fig. S7 [60]), suggesting their potential to host altermagnetism [violet block, Fig. 2(a)]. This number further reduces to eight (as listed in Table I) when considering dynamical stability, as verified by their theoretical phonon spectra (Fig. S8 [60]), which show no significant imaginary frequencies [green block, Fig. 2(a)]. Notably, all eight dynamically stable monolayers originate from prototype-3 [Fig. 1(d)]. In contrast, the remaining 42 monolayers show significant imaginary frequencies in their phonon spectra, indicating dynamical instability in their freestanding form.

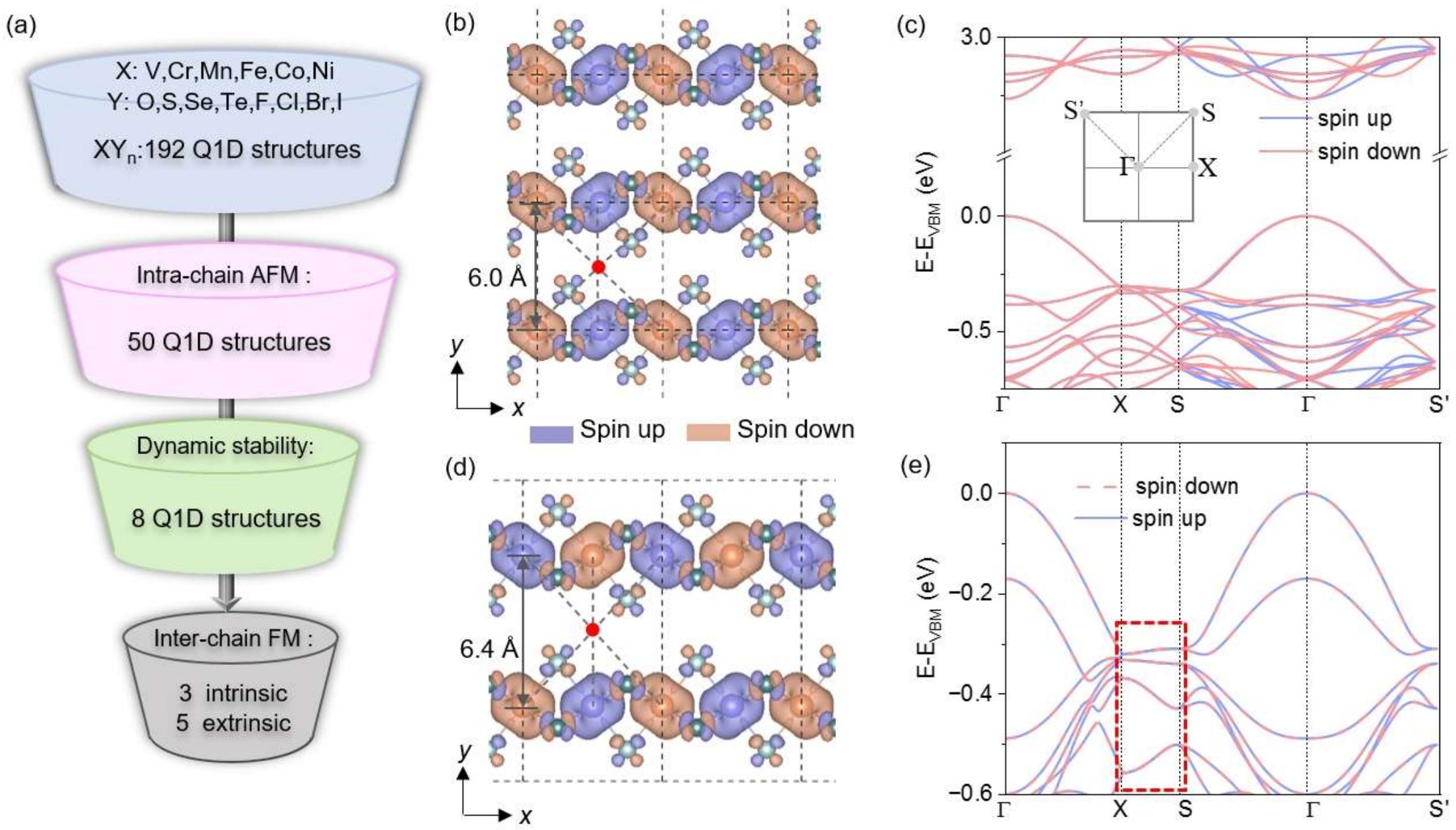

FIG. 2. (a) The screening process of Q1D altermagnets. (b) Top view of spin density distribution and (c) band structure of the $CrCl_3$ monolayer at the inter-chain spacing of 6.0 Å. The red dot represents the inversion center. The illustration shows the high-symmetry path in the Brillouin zone. (d,e) the same scheme of (b,c) for the $CrCl_3$ monolayer with an expanded inter-chain spacing of 6.40 Å. The red dashed box highlights nodal-line electronic states.

Inter-chain FM coupling is presumed in these eight monolayers to host altermagnetism. To clarify the inter-chain magnetic coupling, we used an Ising model Hamiltonian $H = H_0 - \left(\frac{J_1}{2}\sum_{ij} S_i \cdot S_j + \frac{J_2}{2}\sum_{\langle ij\rangle} S_i \cdot S_j + \frac{J_3}{2}\sum_{\langle\langle ij\rangle\rangle} S_i \cdot S_j\right)$, where $ij$, $\langle ij\rangle$ and $\langle\langle ij\rangle\rangle$ represent the nearest, second nearest and third nearest neighboring sites of magnetic atoms, respectively. $S_i$ represents spin operator at magnetic lattice site $i$. Here, $J_1$ to $J_3$ denote spin-exchange parameters illustrated in [Fig. 1(e)], among which $J_3$ specifically represents the inter-chain spin-exchange coupling. Table I summarizes these parameters for the eight Q1D monolayers. Among them, $CrBr_3$, $VCl_3$ and $MnBr_3$ (Table I) exhibit inter-chain FM coupling (positive $J_3$) at their equilibrium inter-chain distances, indicating that their free-standing monolayers are intrinsic altermagnets [gray block, Fig. 2(a)]. Electronic bandstructures, obtained from DFT calculations, further verify their altermagnetic characteristics, displaying significant spin splitting along path S-Γ-S' (Fig. S9 [60]), consistent with our symmetry analysis. However, $CrF_3$ and $CrCl_3$ have near-zero $J_3$ values, resulting in nearly degenerate inter-AFM and FM states. The remaining three Q1D monolayers ($CrI_3$, $FeCl_3$ and $CoTe_3$) exhibit AFM inter-chain coupling, which prevents them from showing altermagnetic characteristics in their free-standing forms.

Table 1. Lattice constants ($a$ and $b$) and spin-exchange parameters ($J_1$, $J_2$, $J_3$, labeled in Fig. 1e, in unit of meV per magnetic atom) of the eight dynamically stable AA-stacked intra-chain AFM $\beta$-$XY_3$-Q1D monolayers.

| Monolayer | $a$ (Å) | $b$ (Å) | $J_1$ (meV) | $J_2$ (meV) | **$J_3$ (meV)** | Altermagnet |
|---|---|---|---|---|---|---|
| $CrBr_3$ | 6.32 | 6.47 | -5.83 | 0.11 | **0.01** | intrinsic |
| $VBr_3$ | 6.30 | 6.59 | -21.66 | 5.53 | **0.01** | intrinsic |
| $MnBr_3$ | 6.43 | 6.39 | -1.13 | 0.43 | **0.48** | intrinsic |
| $CrF_3$ | 5.35 | 4.83 | -6.61 | 0.01 | **0.00** | extrinsic |
| $CrCl_3$ | 5.92 | 6.18 | -10.40 | -0.01 | **0.00** | extrinsic |
| $CrI_3$ | 6.72 | 7.12 | -3.63 | 0.19 | **-0.05** | extrinsic |
| $FeCl_3$ | 5.98 | 6.28 | -1.76 | -0.57 | **-0.19** | extrinsic |
| $CoTe_3$ | 7.75 | 4.74 | -6.95 | -2.26 | **-0.68** | extrinsic |

Among these eight monolayers, three favor the inter-chain AFM coupling. To further assess their dynamic stability, we examined vibrational frequencies of inter-chain AFM coupled structures. The absence of significant imaginary frequencies at the Γ point, along with the non-imaginary frequencies observed in their inter-chain FM coupled counterparts, demonstrates the feasibility of tuning inter-chain magnetic coupling between AFM and FM in these monolayers. The $CrCl_3$ Q1D monolayer, composed of experimentally synthesizable $CrCl_3$ single-atomic chains, exhibits nearly degenerate inter-chain FM and AFM couplings in its freestanding form. Modulating the inter-chain spacing at vdW gaps effectively tunes the inter-chain magnetic coupling [65], which plays a paramount role in the emergence of altermagnetic characteristics. A lower inter-chain force constant facilitates this tuning process. Notably, the inter-chain force constant of $5.06\times10^{19}$ N/m$^3$ for monolayer $CrCl_3$ is only half to the interlayer value for black phosphorus ($10.1\times10^{19}$ N/m$^3$) [66]. This soft inter-chain modulus enables feasible tuning of the inter-chain spacing from the equilibrium lattice constant $b$, 6.18 Å theoretically.

When the spacing is reduced to 6.00 Å, $J_3$ increases to 0.02 meV/Cr, favoring an altermagnetic state. At this distance, the spin density distribution of the $CrCl_3$ monolayer illustrates inter-chain aligned local magnetic moments (the FM state) [Fig. 2(b)]. Although it possesses inversion symmetry, the P-T symmetry has been broken. Band structure calculations verify the expected spin splitting along the S-Γ-S' path [Fig. 2(c)]. On the other side, as the inter-chain spacing expands to 6.40 Å, the nearly zero $J_3$ becomes negative and reaches -0.01 meV/Cr, stabilizing an inter-chain AFM state [Fig. 2(d)]. The two sublattices preserve the inversion and P-T symmetries simultaneously. This configuration corresponds to a Néel AFM semiconductor with a bandgap of 2.84 eV. Our band structure calculations further reveal symmetry-protected fourfold degenerate nodal lines along the X-S direction [Fig. 2(e)] [67]. These results illustrate the tunability of inter-chain spacing for introducing transitions between an altermagnetic semiconductor and a Néel antiferromagnetic semiconductor.

Two distinct trends emerge in magnetic-spacing relations among these eight monolayers. One such trend, exemplified by the $VBr_3$ monolayer, indicates that decreasing inter-chain spacing (lattice constant ***b***) favors the FM coupling [Trend-D, Fig. 3(a)]. This behavior is also observed in $CrBr_3$, $CrI_3$, $CrCl_3$ and $MnBr_3$ monolayers (Fig. S10 [60]). When the inter-chain spacing of the $VBr_3$ monolayer increases from its equilibrium value of 6.59 Å to 6.80 Å [red pentagram in Fig. 3(a)], the inter-chain magnetism becomes AFM from the original FM. The AFM coupling leads to a transition of the monolayer from an altermagnet to a nodal-line semiconductor with a bandgap of 1.65 eV, as shown in Fig. 3(b).

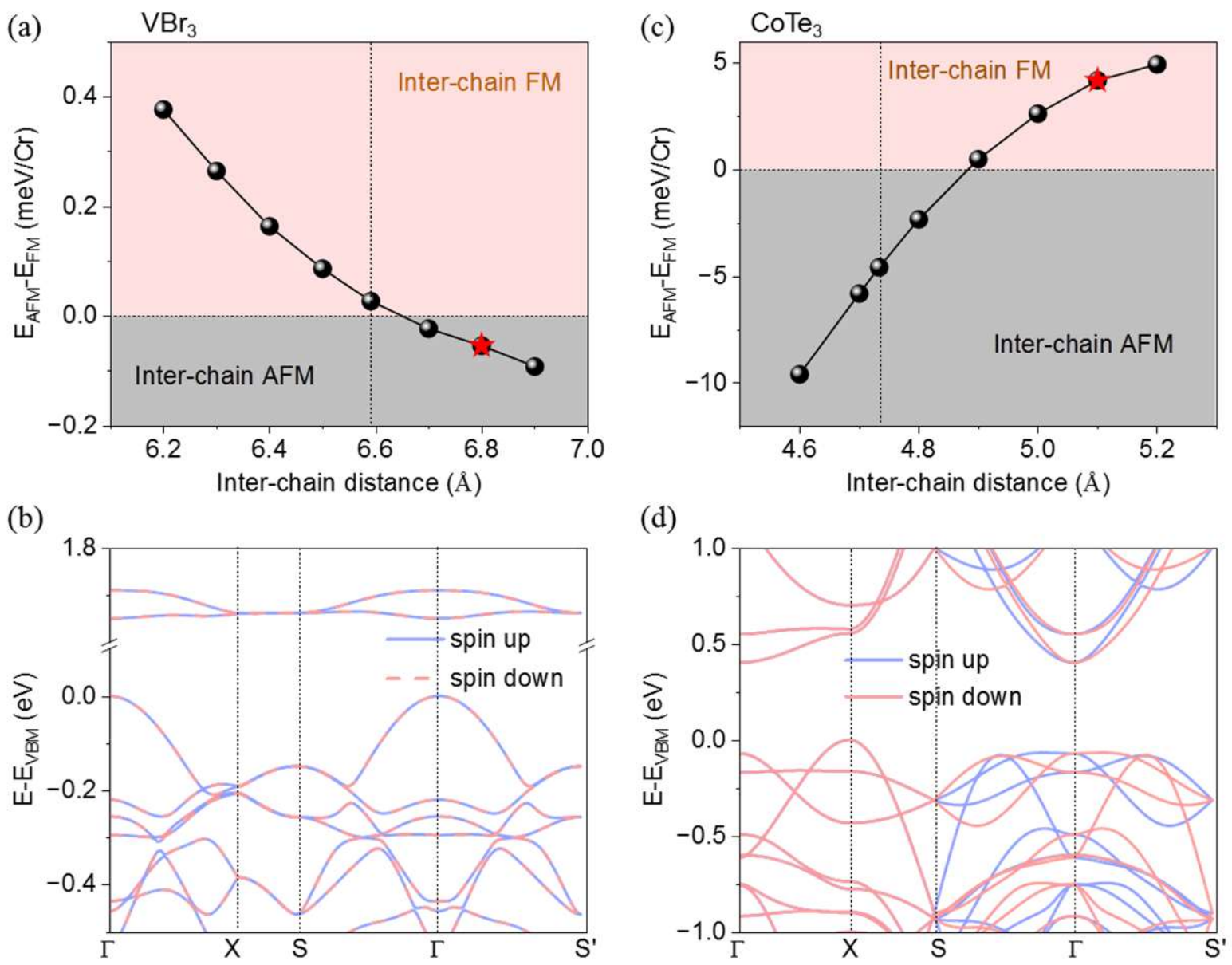

FIG. 3. (a) The energy difference ($E_{AFM}$-$E_{FM}$) as a function of inter-chain spacing for Q1D $VBr_3$ monolayer. The vertical dashed line indicates the freestanding inter-chain distance. (b) Band structure of the monolayer $VBr_3$ under inter-chain of 6.80 Å [labeled as red pentagram in 3(a)]. (c) The energy difference as a function of inter-chain spacing for Q1D $CoTe_3$ monolayer. (d) Band structure of the monolayer $CoTe_3$ under inter-chain of 5.10 Å [labeled as red pentagram in 3(c)].

The $CoTe_3$ monolayer follows an opposite trend, where increasing the inter-chain spacing prefers the inter-chain FM coupling (Trend-I). As shown in Fig. 3(c), a gradual increase in the inter-chain spacing transitions the initially favored inter-chain AFM coupling to FM. The inter-chain spin-exchange parameter $J_3$ reduces from -0.68 meV/Co to zero and subsequently increases to 1.09 meV/Co (Table I) as the inter-chain spacing expands from its equilibrium (4.74 Å) to 5.10 Å [red pentagram in Fig. 3(c)]. At the 5.10 Å spacing, significant spin splitting is observed in the valence band along the S-Γ-S' path [Fig. 3(d)]. This trend, where increasing the inter-chain spacing weakens the inter-chain AFM coupling, is also consistently observed in $CrF_3$ and $FeCl_3$ monolayers (Fig. S11 [60]).

The inter-chain magnetic ground state is determined by the competition between Pauli repulsion and inter-chain hopping [65,68]. Unlike the 2D case, Q1D monolayers exhibit two distinct inter-chain bonding states that result in Pauli repulsion at vdW gaps, resulting in double antiferromagnetic regions in the Bathe-Slater curve (Fig. S12 [59]). This two-orbital Bathe-Slater curve features two Trend-I regions and one Trend-D region. The equilibrium inter-chain distances of $CrCl_3$, $VBr_3$, $CrI_3$, $CrBr_3$, and $MnBr_3$ fall within the Trend-D region, while those of $CrF_3$, $FeCl_3$, and $CoTe_3$ lie in the Trend-I region. Among these eight Q1D monolayers, $CoTe_3$ exhibits the largest spin-splitting, reaching several hundred meV. While $CrF_3$ and $MnBr_3$ show the largest (4.56 eV) and smallest (0.19 eV) bandgaps, respectively (Table S1). Recently experimental studies demonstrate that hydrostatic pressure [69], nanotube radius variation [70], and coverage adjustment [71] can tune inter-chain distances in 1D systems. We also infer that substrate-mediated tuning can be achieved via vdW epitaxy. These experimental achievements, together with those results discussed earlier, establish inter-chain spacing as an effective means for tailoring electronic and magnetic properties in Q1D monolayer altermagnets.

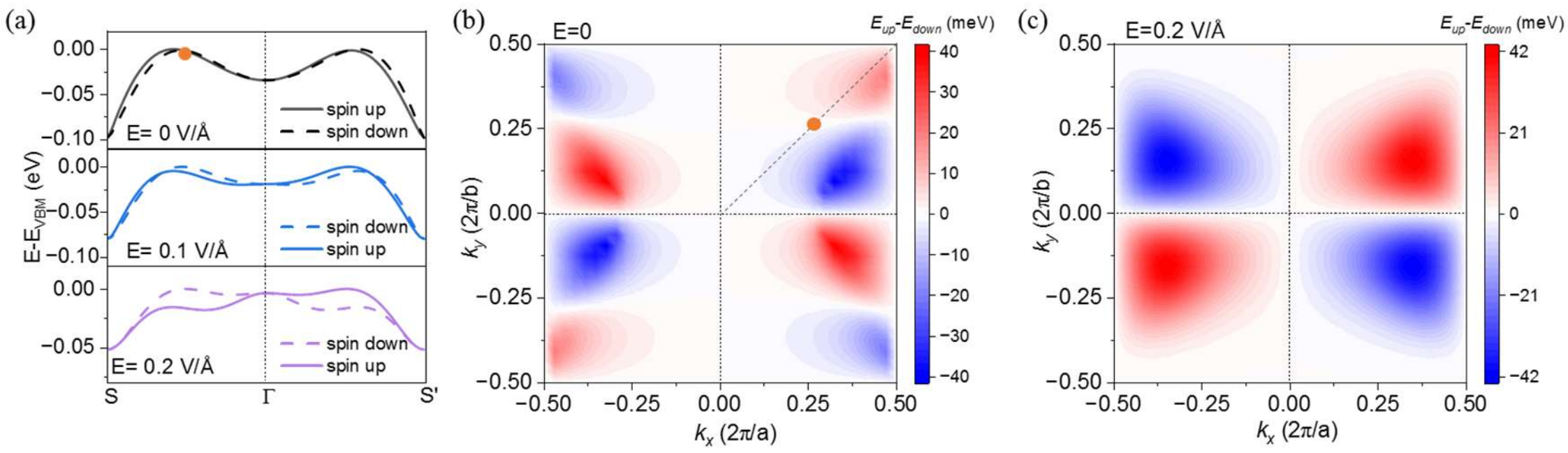

FIG. 4. (a) Band dispersion plots of the highest valence band in freestanding $CrCl_3$ monolayer with inter-chain FM coupling under varied external electric field. The orange dots indicate the band crossing point along the Γ-S direction. Spin splitting mappings of the highest valence band in the freestanding $CrCl_3$ monolayer (b) without electric field and (c) under an electric field of 0.2 V/Å.

We next examined the effect of electric fields on altermagnetism, using monolayer $CrCl_3$ as a representative. Due to its mirror symmetry along the $z$-axis, monolayer $CrCl_3$ exhibits identical response behaviors under both positive and negative electric fields.

Therefore, we focus on the effects of a positive electric field in the following discussion. In the inter-chain FM configuration, the valence bands of the freestanding $CrCl_3$ monolayer, mainly composed of Cl-$p$ orbitals (Fig. S13 [60]), show relatively weak spin splitting (~10 meV) along the S-Γ-S' path. The splitting nearly doubles (~20 meV) at an electric field of 0.2 V/Å [Fig. 4(a)]. An accidental band crossing between spin-up and spin-down components occurs in the valence band at 0 V/Å, which was captured by the bandstructure plot along path Γ-S [orange dot in Fig. 4(a) and 4(b)]. This crossing leads to spin-component changes, and when combined with the preserved $C_{2x}$ rotation and $M_x$ mirror symmetries, the map exhibits an eightfold pattern in the momentum-resolved spin splitting distribution. Upon applying an out-of-plane electric field of 0.2 V/Å, the broken inversion symmetry enlarges the spin splitting, which avoids the accidental crossing spanning the first Brillouin zone, thereby reducing the spin splitting pattern from eightfold to fourfold.

For the inter-chain AFM configuration of the $CrCl_3$ monolayer, while the electric field breaks the P-T symmetry, a fractional translational symmetry of (0, 1/2, 0) prevents the emergence of altermagnetism. However, in certain structures, inter-chain sliding [$CrF_3$, Fig. S14(a) [60]] or rotating [$CoTe_3$, Fig. S14(c) [60]] occurs in inter-chain AFM configurations (Fig. S14 [60]). In these cases, an external electric field breaks both the P-T and the fractional translational symmetries, inducing spin splitting. Although the magnitude of spin splitting in these structures is relatively small (less than 2 meV), this finding demonstrates the rich tunability inherent in Q1D monolayers. These results highlight the flexibility and tunability of altermagnetism in Q1D monolayers under an external electric field.

In summary, we extended the family of altermagnetic materials to Q1D single-atomic magnetic chain structures using a self-assembly strategy. Through symmetry analysis and high-throughput calculations, we predicted eight stable Q1D monolayer, including three intrinsic and five tunable altermagnets, assembled from AA-stacked and intra-chain AFM coupled $\beta$-$XY_3$ single-atomic magnetic chains. These monolayers have highly anisotropic inter-atomic interactions (covalent versus vdW interactions) and electronic structures. They also exhibit various band gaps, and tunable spin splitting. Our study highlights the feasible tunability and the potential for artificial engineering of altermagnetism through inter-chain spacing in these highly anisotropic Q1D monolayers assembled from single-atomic chains. Specifically, we identified two distinct trends. In Trend-I, increasing inter-chain spacing stabilizes FM inter-chain coupling in monolayers like $CoTe_3$, leading to enhanced spin splitting. In Trend-D, a decrease in inter-chain spacing strengthens altermagnetism in monolayers like $VBr_3$. Moreover, external electric fields provide an additional degree of control over spin splitting in altermagnetism, demonstrating the flexibility of Q1D altermagnets for potential device integration. While our study establishes the dynamical stability of AA-stacked $\beta$-$XY_3$ altermagnets, their thermodynamic stability warrants further investigations [76]. Additionally, although free-standing $\gamma$-$XY_2$ monolayers are dynamically unstable, stabilization strategies, such as substrate engineering or charge doping, present promising avenues for exploration. This work broadens the scope of low-dimensional altermagnets and provides a blueprint for designing novel spintronic materials with tunable magnetic and electronic properties.

## Acknowledgements

We gratefully acknowledge the financial support from the National Natural Science Foundation of China (Grants No. 92477205 and No. 52461160327), the National Key R&D Program of China (Grant No. 2023YFA1406500, 2021YFA1401500), the Hong Kong Research Grants Council (16306220, 16303821, 16306722 and 16304523), the Fundamental Research Funds for the Central Universities, and the Research Funds of Renmin University of China (Grants No. 22XNKJ30). Calculations were performed at the Hefei Advanced Computing Center, the Physics Lab of High-Performance Computing (PLHPC) and the Public Computing Cloud (PCC) of Renmin University of China.

## DATA AVAILABILITY

The data that support the findings of this article are not publicly available upon publication because it is not technically feasible and/or the cost of preparing, depositing, and hosting the data would be prohibitive within the terms of this research project. The data are available from the authors upon reasonable request.